# MASTER prompt and follow-up GRB observations

**Nataly Tyurina[1], Vladimir Lipunov[1], Victor Kornilov[1], Evgeny Gorbovskoy[1], Nikolaj Shatskij[1], Dmitry Kuvshinov[1], Pavel Balanutsa[1], Alexander Belinski[1], Vadim Krushinsky[2], Ivan Zalozhnyh[2], Andrey Tlatov[3], Alexander Parkhomenko[3], Kirill Ivanov[4], Sergey Yazev[4], Petr Kortunov[1], Anatoly Sankovich[1], Artem Kuznetsov[1]**

[1] *Moscow State University, Sternberg Astronomical Institute, 119991, 13, Univeristetskij pr-t, Moscow, Russia*

[2] *Ural State University, 620083, 51, Lenina pr-t, Ekaterinburg, Russia*

[3] *Kislovodsk Solar Station, 357700 p.o. Box 145, 100, Gagarina st., Russia*

[4] *Irkutsk State University, 664003, 1, Karl Marks ul., Irkutsk, Russia*

There are the results of gamma-ray bursts observations obtained using the MASTER robotic telescope in 2007–2009. We observed 20 error-boxes of gamma-ray bursts this period.The limits on their optical brightnesses have been derived. There are 5 prompt observations among them, obtained at our very wide field cameras. Also we present the results of the earliest observations of the optical emission of the gamma-ray bursts GRB 050824 and GRB 060926.

## 1. Introduction

The construction of robotic telescopes, which not only automatically acquire but also automatically process images and choose observing strategies, is rather new and vigorously developing area in modern astronomy. MASTER (Mobile Astronomy System of Telescope Robots, http://observ.pereplet.ru) is the first and still unique robotic telescope system in Russia, began to be created through the efforts of scientists at the Sternberg Astronomical Institute of Moscow State University and the Moscow "Optika" Association in 2002, and continues to be developed to the present MASTER-NET (Lipunov [1, 2, 3]).

MASTER is dedicated to observation and detection of optical transients on time scales of seconds to days. The emphasis is on gamma-ray bursts (GRBs), the most powerful explosions in our Universe. A special program package for image reduction in real time has been created,making it possible not only to carry out astrometry and photometry of a frame, but to recognize objects not contained in astronomical catalogs: supernovae, new asteroids, optical transients, and so forth. This software let us open several supernovae SN2008gy, SN 2006ak, SN2005ee, SN 2005bv).

We observed 20 error boxes of gamma-ray bursts in 2007-2009 y.y. The results of them are presented. There are 5 prompt observations, obtained by our very wide field cameras (VWFC) during last half-year after we mount these cameras in Kislovodsk and Irkutsk in autumn 2009.

## 2. MASTER system

Now MASTER system has 4 bases: near Moscow, near Kislovodsk, near Irkutsk and near Ekaterinburg (Lipunov [3]). We have far observatories only a year and a half, and of about 5 years we observed only from Moscow telescope system, that are in village near Domodedovo airport. This small distances from Moscow State University let us design, test and modify robot system, repair possible technical fault and write reduction software without state financial support. But minus is very bad weather condition near megapolis (less than 70 nights per yr and most of them are in summer white nights)). But we observed several tens error boxes during 5 years (HETE and Swift epochs) and in most of them our observations were the earliest (Lipunov [2]).

MASTER characteristics (Lipunov [3]) are closest to the american ROTSE-III system [4] (http://www.rotse.net). MASTER differs in its larger field of view and the presence of several telescopes on a single axis, which makes us possible to obtain images at several different wavelengths simultaneously, that was realized in MASTER-Ural observations.

So as up to 2008 year most of GRB observations were at MASTER-Moscow system, there are the following characters of it.

The main telescope (355 mm diameter) takes images in white light, and is the main search element of the system. An Apogee Alta U16 (4096 × 4096 pixels) is installed on it, making it possible to obtain images in a six square degree field. In addition, MASTER has a very-wide-field camera ($50^{\circ} \times 60^{\circ}$) that covered the field of view of the HETE orbiting gamma-ray telescope, made it possible to obtain simultaneous observations with HETE to $9^{m}$ using a separate automated scheme. This widefield equipment enables searches only for bright, transient objects.

Now we have very wide field cameras at Kislovodsk on the Mountain Solar Station of the Main Astronomical Observatory in Pulkovo, making it possible to continuously monitor a 420-square-degree field of sky to 13*m* in a five second exposure. Such cameras are now in Irkutsk.

The Kislovodsk, Domodedovo, Ural and Irkutsk systems are connected via the Internet, and are able to respond to the detection of uncataloged objects (optical transients) within several tens of seconds (including processing time). The results of observations using the MASTER network will be reported separately.

All MASTER system are able to operate in a fully automated regime: automatically, based on the ephemeredes (sunset) and the presence of satisfactory weather conditions (the control computer is continuously attached to a weather sensor), the roofs (above the main mount and wide-field camera) are opened, the telescopes are pointed at bright stars and pointing corrections introduced, and, depending on the seeing, it then either goes into a standby regime or begins a survey of the sky using a specialized, fully automated programme.

Thus, observations are conducted in two regimes: survey and “alert” (e.g. observations of the locations of gamma-ray bursts based on coordinates obtained). In the former case, the main telescopes automatically takes three frames of an arbitrary region in succession, with exposures from 30 to 60 s, moves to a neighboring region $2^{\circ}$ away and carries out the same procedure, and so on, repeating a given set of three frames every 40–50 min. This makes it possible to avoid artefacts in the data processing, and to locate moving objects.

The alert regime is supported by a continuous connection between the control computer and the GCN international gamma-ray burst (GRB) network (http://gcn.gsfc.nasa.gov). After detection of a GRB by a space gamma-ray observatory SWIFT, Konus-Wind, FERMI etc.), the telescope obtains the coordinates of the burst region (coordinate error box), automatically points to this direction, obtains an image of this region, reduces this image, and identifies all objects not present in the computer catalogs. If a GRB is detected during the day, its coordinates are included in the observing program for the next night.

A special program package for image reduction in real time has been created,making it possible not only to carry out astrometry and photometry of a frame, but to recognize objects not contained in astronomical catalogs: supernovae, new asteroids, optical transients, and so forth. Over the entire time observations have been obtained on the MASTER system (see the results of 2002-2006 observations in [1, 2, 5]), images have been obtained for almost 100 GRB error boxes at this moment. In a half cases, these observations were the first in the world.

### 3. GRB observations

During 2007-2009 years we observed 20 GRB. The results of our prompt and follow-up observations are in Table 1.

**Table 1.** Table caption: GRB observations by MASTER system in 2007-2009y.y..

| Burst | Optical limits,m | Comments (observatory, time after trigger information, exposures, circular number, etc.) |
|---|---|---|
| GRB090528B | R>19.0<br>V>18.1 | MASTER-URAL, 7h after GRB, 180-s exposures in R and V filters. Our images cover 30% last 1-σ Fermi error box. No OT. (Lipunov [6]) |
| **GRB090424** | ~8, ~9 | **Prompt observations** by 6 MASTER Very Wide Field Cameras (VWFC) in Kislovodsk and Irkutsk with common FOW = 6000 sq.deg., 1-s exposures. Unfiltered images (very close to V band). No OT. Lipunov [7] |
| GRB090408B | 17 | MASTER-URAL , R and V 60-s exposure, starting after sunset (87min after trigger time). No OT. GCN9111 (Lipunov [8]) |
| GRB090328B | 9.1 | **Prompt observation** at MASTER VWFC located at Irkutsk. 1s exposures during night: before, during and after GRB. No OT. GCN9065 (Lipunov [9]) |
| GRB090320B | V ~9.0 | **Prompt observation** at MASTER VWFC located at Kislovodsk. We observed ~80% 1-σ Fermi error box with 1-s exposure during all night: before, during and after GRB-time. No OT. GCN9038 (Lipunov [10]) |
| GRB090305B | V~9.5 | **Prompt observation.** MASTER VWFC located at Kislovodsk observed ~80% 1-σ Fermi error box with 1-s exposure 7 hours before, during and 1 hours after GRB Time .GCN9004 (Lipunov [11]) |
| GRB081215 A | V~11 | MASTER VWFC located at Kislovodsk observed Fermi trigger (SGR 0044+42) during all night with 5-s expositions. GCN8674,8673,8672 (Lipunov [12]) |
| GRB081130B | V~12 | **Prompt observation.** MASTER VWFC located at Kislovodsk observed Fermi trigger with 5s exposures(each emage is V~11$^m$) No OT. GCN8597,8585 (Lipunov [13]) |
| GRB081110 | 19 | MASTER-Moscow observed error box 6 hours 42 min after the GRB time. We have 89 images with limit for coadded one up to 19$^m$ (S/N = 4). No OT. GCN8518 (Lipunov [14]) |
| GRB081102 | V~13 | 2 of MASTER VWFC at Kislovodsk observed this error box with 5s exposures during all night. There are two separeted (~702m) mount with double cameras. GCN8516,8471,8464 (Lipunov [15]) |
| GRB080822B | 18.8 | MASTER-Moscow observed it 18m 43s after the GRB time with 30-s exposures. Each image has 17$^m$, Summary image (25x30s) has 18.8$^m$. The unfiltered images are calibrated relative to USNO A2.0 (0.8 R + 0.2 B). GCN8123 (Lipunov [16]) |
| GRB080605 | R~11.5 | MASTER VWFC at Kislovodsk observed it 46s after the GRB-time (12 s after the notice arrival time) with a series of 5s exposures. Unfiltered (close to V). No OT. GCN7836(Lipunov [17]) |
| GRB080319 D | V~11.5 | MASTER VWFC at Kislovodsk observed Swift-BAT trigger with a series of 5s exposures starting 92s after notice arrival time (708s after GRB time). Unfiltered (close to R, i.e. another camera). No OT. GCN7455 (Lipunov [18]) |
| GRB080205 | 19.5 | MASTER-Moscow observed it 481 min after the GRB-time. No OT. GCN7261 (Lipunov [19]) |
| GRB071122 | 16 | MASTER-Moscow observed it 151 s after the GRB time (61s after notice-time). No OT. GCN7129 (Lipunov [20]) |
| GRB070810. 8 | 14.8 | MASTER -Moscow observed it 125s after the GRB time. No OT. GCN 6752 (Lipunov [21]) |
| GRB070224 | 13, | MASTER VWFC observed error box in Kislovodsk with series of 5s exposures starting 2 s after notice arrivel time. Unfiltered (close to R). No OT. |
| | 18 | MASTER-Moscow observed error box 51s after notice arrivel time. Unfiltered. No OT. GCN6140,6139,6138 (Lipunov [22]) |
| GRB070223 | 13 | MASTER VWFC observed error box in Kislovodsk 5s exposures starting 1 s after notice arrivel time. No OT. |
| | | Unfiltered (close to R). GCN6131 (Lipunov [23]) |
| GRB070219 | 13.5 | MASTER VWFC observed error box in Kislovodsk 5s exposures starting 76s after the GRB time and 15 s after notice arrivel. No OT. GCN6113 (Lipunov [24]) |

Kislovodsk and Irkutsk VWFC had different cameras during 2007 and 2008 year, so there are V and R bands in second column of the table. We have unfiltered images from these cameras, but they are closest to V and R bands, and we calibrated using TICHO catalogue (Gorbovskoy, [25]).

There were **5 prompt** observations by very wide field cameras in Irkutsk and Kislovodsk stations during last half-year (after mounting several VWFC there): GRB090424, GRB090328B, GRB090320B, GRB090305B, GRB081130B. There were no optical candidate found at reducted images and we give the optical limits for GRBs.

We also calculate the ratio of optical to gamma fluence (Gorbovskoy [25]) for GRB.

There were several interesting GRB in earliest year, let us discuss them shortly.

**GRB 050824**

MASTER have the earliest images of this GRB. Information about the coordinates of GRB 050824 arrived at the MASTER observatory on August 24, 2005 after some delay due to the processing of the signal in the SWIFT data center. The first image of the region was obtained 110 s seconds after obtaining the alert, i.e., 764 s after the SWIFT detection (trigger 151905), during a nearly fully Moon. The optical limit reached was $17.8^m$. The earliest available optical image of this object, obtained on our MASTER system, can be found at the address http://observ.pereplet.ru/images/GRB050824/1.jpg. Figure1 presents the upper imits and magnitudes obtained in the first minutes of observations. Figure 2 shows our data together with the *R* data of the MDM observatory. The MDM *R* observations obtained from 5.6 to 12.6 h after the GRB are consistent with a power-law decay in the flux (with index −0.55 ± 0.05), and also with our data obtained 24 min and 47 min after the burst.

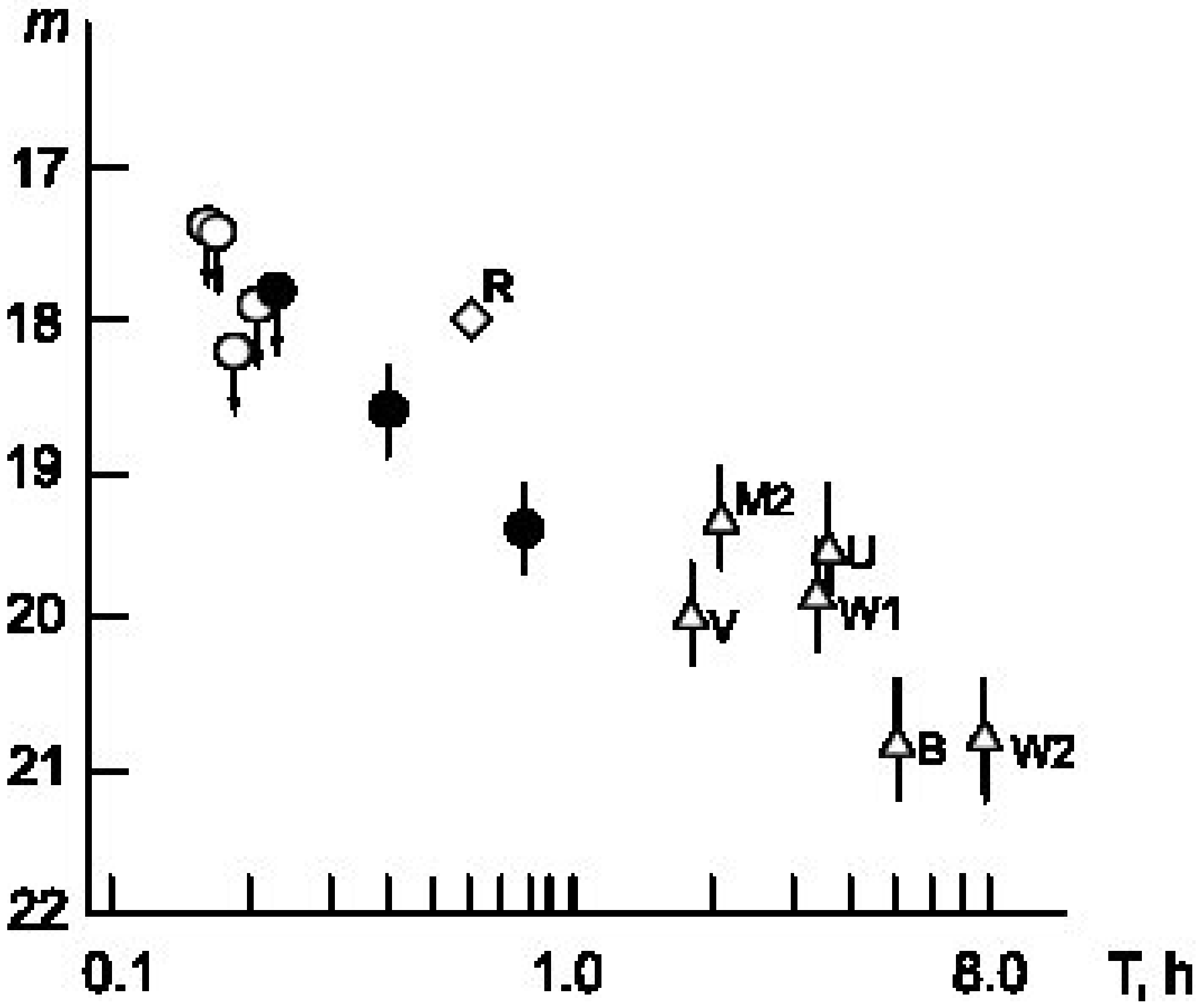

Figure 1. ROTSE-III (Akerlof, [26]) (opened circles), MASTER (Lipunov, [27]) (dark circles), OSN telescope (J.Gorosabel, [28]) (romb) and SWIFT (P. Schady, [29]) (triangles) observations of GRB050824. Axies are magnitudes and times in hours.

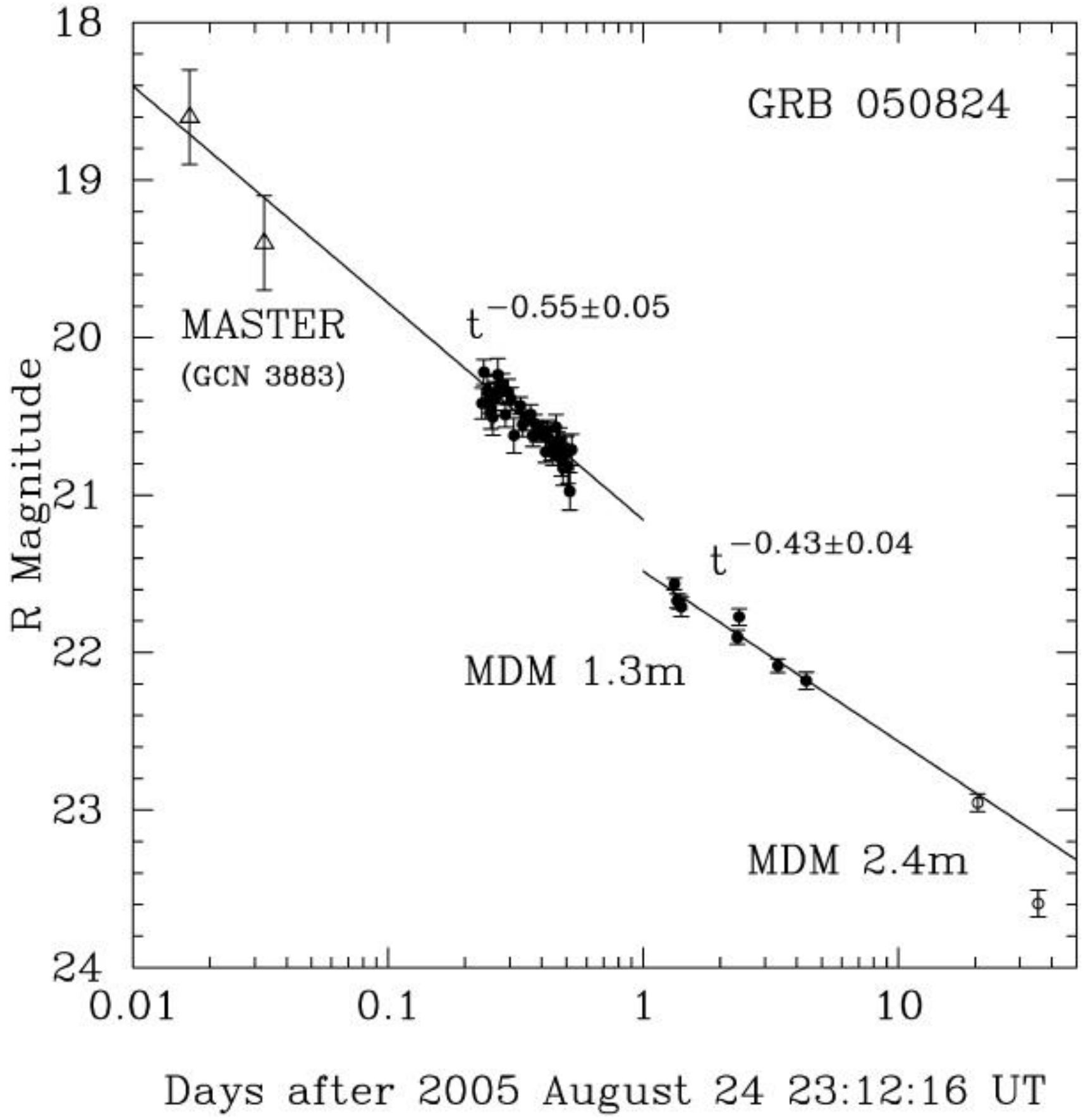

**Figure 2.** The power decay for GRB050824, using MASTER (Lipunov, [27]) and MDM (Halpern, [30]) observations.

**GRB 051103**

GRB 051103 may be the first soft gamma-ray repeater (SGR) detected outside our Galaxy. The first image of the error-box region was obtained on the MASTER telescope (Lipunov [5]). The bright, short (0.17 s) burst GRB 051103 was detected by Konus-Wind, as well as HETE-Fregate, Mars Odyssey (GRS and HEND), RHESSI, and SWIFT-BAT. The MASTER telescope started to observe the error-box region for GRB 051103 several minutes after receiving the alert telegram. We obtained 36 images with a total exposure time of 1080s. No optical transient was found in the error box to $18.5^m$ (in the presence of a fullMoon and light haze). Analysis of the frame showed the presence of four bright galaxies near or in the large error box. The most likely candidate host galaxy is M81, and the burst itself has been interpreted as a SGR. The error box lies outside any spiral arms, where strongly magnetized neutron stars (magnetars, which are thought to be the sources of SGRs) would be likely to form. However, the structure of M81 is distorted by tidal interaction, and part of a disrupted spiral may fall in the error box. For example, the ultraluminous X-ray source (ULX) M81 X-9 is located at a similar distance from the center of M81 (on the side opposite to the error box), and belongs to that galaxy's population of massive stars. It would be interesting to search for supernova remnants within the error box (unfortunately, the supernova-remnant survey does not encompass the error box of the GRB). We should also noted the elliptical galaxy PGC 028505, which is close to the center of the triangulation error box.

Our full image of the error box region can be found at the address http://observ.pereplet.ru/images/GRB051103.4/sum36.jpg.

**GRB 060926**

Our observations of GRB 060926, detected by the SWIFT gamma-ray observatory, were carried out in an automated regime under good weather conditions. The earliest image was obtained on 76s after the detection of the burst. We found an optical transient in our first and subsequent summed frames, which coincides within the errors with the coordinates of the optical transient. Our photometry of the object provided the earliest points on the light curve (Lipunov [5]).

Our preliminary reduction indicated a more gradual brightness decrease than the OPTIMA-Burst observations (the power-law index for the brightness decrease in the first 10 min was 0.69). However, subjecting the data to finer time binning revealed an optical burst: after its initial decrease, the brightness began to increase beginning 300 s after the GRB, reaching a maximum 500–700 s after the GRB. Synchronous SWIFT-XRTmeasurements of the X-ray flux show similar behavior .

Such an event had already been observed in at least two other cases: GRB060218A ($z$ = 0.03) 1000 s after the GRB, and GRB060729 ($z$ = 0.54) 450 s after the GRB.
The absorption indicated by the X-ray data corresponds to $n_H = 2.2 \times 10^{21}$ cm$^{-1}$, of which $n_H = 7 \times 10^{20}$ cm$^{-2}$ occurs in the Galaxy. Taking into account the redshift, the total absorption in our band should be $\approx 3^m$. Naturally, we assume here that the dust-to-hydrogen ratio is the same as it is in our Galaxy. A comparison of our optical data with the SWIFTXRT X-ray fluxes can be used to determine the slope $\beta$ of the electromagnetic spectrum, which turned out to be constant within the errors and equal to $1.0 \pm 0.2$, which coincides with the corresponding value for the X-ray spectrum.

The earliest image of the optical transient can be found at http://observ.pereplet.ru/images/GRB060926/GRB060926_1.jpg, and the sum of the ten following frames at http://observ.pereplet.ru/images/GRB060926/GRB060926_10.jpg.

## 4. Conclusions

We presented the results of last years GRB observations obtained on the MASTER robotic telescope, which is the only telescope of its kind in Russia. These results include 5 prompt observations of GRB in 2008 and 2009yy., follow-up observations of 15 another GRB in 2007-2009, and the first observations of optical emission from the gamma-ray bursts GRB 050824 and GRB 060926. Our data together with observations made later yield a brightness-variation law for GRB 050824 of $t^{-0.55\pm0.05}$.

Our experience of two years of operation of the MASTER wide-field robotic telescope has demonstrated its unique capabilities. We hope, that new MASTER telescopes at Kislovodsk and Ural stations, and also new Irkutsk and Blagoveschwnsk systems, that are made now, will justify our hopes. If such systems could be installed at suitable sites at various hour angles across Russia, they would provide unique information via continuous monitoring of both the near and distant cosmos.

**Ackowledgment**

The authors thank the General Director of the “OPTIKA” Association S.M. Bodrov for providing the MASTER project with necessary expensive equipment.